\documentclass[aps,prd,groupedaddress,showpacs]{revtex4}

\usepackage{graphicx,dcolumn,bm,amssymb,amsmath,latexsym,amsfonts,footnote}

\begin{document}


\newcommand{\nonu}{\nonumber}
\newcommand{\sm}{\small}
\newcommand{\noi}{\noindent}
\newcommand{\npg}{\newpage}
\newcommand{\nl}{\newline}
\newcommand{\bp}{\begin{picture}}
\newcommand{\ep}{\end{picture}}
\newcommand{\bc}{\begin{center}}
\newcommand{\ec}{\end{center}}
\newcommand{\be}{\begin{equation}}
\newcommand{\ee}{\end{equation}}
\newcommand{\beal}{\begin{align}}
\newcommand{\eeal}{\end{align}}
\newcommand{\bea}{\begin{eqnarray}}
\newcommand{\eea}{\end{eqnarray}}
\newcommand{\bnabla}{\mbox{\boldmath $\nabla$}}
\newcommand{\univec}{\textbf{a}}
\newcommand{\VectorA}{\textbf{A}}
\newcommand{\Pint}

\title{Extreme binary black holes in a physical representation}

\author{I. Cabrera-Munguia\footnote{icabreramunguia@gmail.com}}
\affiliation{ Departamento de F\'isica y Matem\'aticas, Universidad Aut\'onoma de Ciudad Ju\'arez, 32310 Ciudad Ju\'arez, Chihuahua, M\'exico}


\begin{abstract}
Stationary axisymmetric binary systems of unequal co and counter-rotating extreme Kerr black holes apart by a conical singularity are studied. Both solutions are well identified as two $3$-parametric subfamilies of the Kinnersley-Chitre metric, and fully depicted by Komar parameters: the two masses  $M_{1}$ and $M_{2}$, and a coordinate distance $R$, where the angular momenta $J_{1}$ and $J_{2}$ are functions of these parameters. Our physical representation allows us to identify some limits and novel physical properties.
\end{abstract}

\maketitle

\vspace{-0.5cm}
\section{Introduction}
\vspace{-0.5cm}
The well-known Kinnersley-Chitre (KCH) $5$-parametric exact solution \cite{KCH} represents the extreme limit case of the so-called double-Kerr-NUT solution developed by Kramer and Neugebauer in 1980 \cite{KramerNeugebauer}, which allows to treat the superposition of two massive rotating sources in General Relativity. Both solutions permits us to study the dynamical interaction among two Kerr-type sources in stationary axisymmetric spacetimes by solving properly the corresponding axis conditions. In this respect, Yamazaki \cite{Yamazaki} found an asymptotically flat special member of the KCH metric through a specific parametrization that vanishes the NUT parameter \cite{NUT} which is identical to the Tomamitsu-Sato solution with distortion parameter $\delta=2$ \cite{TS}. A few years ago, after following the ideas provided by Yamazaki \cite{Yamazaki} to eliminate the NUT parameter, Manko and Ruiz \cite{MR} solved for the first time in analytical way the axis condition that disconnects the region in between sources, with the main purpose to describe co and counter-rotating binary black hole (BH) systems separated by a conical singularity \cite{Bach,Israel}; i.e, a massless strut related with the interaction force among sources which is a measure of their gravitational attraction as well as the spin-spin interaction. Even though the Manko-Ruiz representation of the KCH metric allows us to clarify some physical aspects related to unequal binary systems, the total Komar \cite{Komar} mass $M$ and total angular momentum $J$ of the binary BH system contain complicated formulas in terms of dimensionless parameters, which could lead to erroneous interpretations at the moment of assigning numerical values to them. Therefore, it is mandatory to review once again the KCH solution in order to express the metric of two-body systems of unequal co and counter-rotating extreme BHs separated by a strut in a representation with a more physical aspect.

The main goal pursued in this paper is a rederivation of the two 3-parametric subfamilies of the KCH metric concerning to co/counter-rotating BHs considered earlier in \cite{MR}, but with the principal characteristic that now both solutions will be given in terms of arbitrary physical Komar parameters: the masses $M_{1}$ and $M_{2}$, as well as the coordinate distance $R$. We will obtain some well-known limits of the KCH solution and other dynamical aspects not considered before; in particular, those related to the merging process of interacting BHs. The paper is organized as follows. In Sec. II we describe the KCH exact solution as well as the two approaches considered earlier in Refs.\ \cite{Yamazaki,MR}; in particular, the path used by Manko and Ruiz to solve the axis conditions in order to describe interacting binary BHs by means of two 3-parametric special members of the KCH metric. Later on, in Sec. III we begin with a new more suitable $5$-parametric representation of the KCH solution with the main objective to solve once again the axis conditions and depict both metrics for interacting BHs in a more realistic physical representation. Concluding remarks can be found in Sec. IV.

\vspace{-0.5cm}
\section{The KCH exact solution}
\vspace{-0.5cm}
Stationary axisymmetric spacetimes are well defined with the Papapetrou metric \cite{Papapetrou}
\be ds^{2}=f^{-1}\left[e^{2\gamma}(d\rho^{2}+dz^{2})+\rho^{2}d\varphi^{2}\right]- f(dt-\omega d\varphi)^{2},\label{Papapetrou}\ee

\noi and Einstein vacuum field equations can be reduced by means of Ernst's formalism \cite{Ernst} into a new complex equation
\be ({\cal{E}}+ \bar{\cal{E}})({\cal{E}}_{\rho \rho} + \rho^{-1}{\cal{E}}_{\rho}+{\cal{E}}_{z z})
=2({\cal{E}}_{\rho}^{2}+{\cal{E}}_{z}^{2}), \label{Ernsteq}\ee

\noi where a suffix $\rho$ or $z$ denotes partial differentiation. It follows that one can be able to find the metric functions $f(\rho,z)$, $\omega(\rho,z)$, and $\gamma(\rho,z)$ of the line element Eq.\ (\ref{Papapetrou}) by solving the following equations:
\bea \begin{split}  f&=  {\rm{Re}}({\cal{E}}), \\
\omega_{\rho} &= -4\rho ({\cal{E}}+ \bar{\cal{E}})^{-2}{\rm{Im}}({\cal{E}}_{z}),\qquad
\omega_{z} = 4\rho ({\cal{E}}+ \bar{\cal{E}})^{-2}{\rm{Im}}({\cal{E}}_{\rho}),\\
\gamma_{\rho}&=\rho ({\cal{E}}+ \bar{\cal{E}})^{-2} \left({\cal{E}}_{\rho} \bar {\cal{E}}_{\rho} -{\cal{E}}_{z} \bar {\cal{E}}_{z}\right),\qquad
\gamma_{z}=2\rho ({\cal{E}}+ \bar{\cal{E}})^{-2} \rm{Re}({\cal{E}}_{\rho}\,{\bar{\cal{E}}}_{z}),
\label{metrics}\end{split}\eea

\noi once we know an analytical solution for the non-linear Eq.\ (\ref{Ernsteq}). In this sense, the KCH solution solves Eq.\ (\ref{Ernsteq}) exactly, it is described by the complex potential ${\cal{E}}$ which is given by$^{[30]}$\footnotetext[30]{Kinnersley and Chitre used the inverse function of ${\cal{E}}$ in their original paper \cite{KCH}, i.e., $\xi=\frac{1-{\cal{E}}}{1+{\cal{E}}}=\frac{2\Gamma}{\Lambda}$. }
\bea \begin{split} {\cal{E}}&=\frac{\Lambda-2\Gamma}{\Lambda+2\Gamma}, \\
\Lambda&=(\alpha^{2}-\beta^{2})(x^{2}-y^{2})^{2}+p^{2}(x^{4}-1) +q^{2}(y^{4}-1)-2i \alpha(x^{2}+y^{2}-2x^{2}y^{2})\\
&-2i p q xy (x^{2}-y^{2})-2i \beta xy(x^{2}+y^{2}-2), \\
\Gamma&=e^{-i\gamma_{o}}[px(x^{2}-1) + iqy(y^{2}-1)-i(p \alpha+iq\beta)x(x^{2}-y^{2}) +i(p \beta + iq \alpha)y(x^{2}-y^{2})], \label{ErnstKCH} \end{split}\eea

\noi where $(x,y)$ are prolate spheroidal coordinates depicted as
\be x=\frac{r_{+}+r_{-}}{2\kappa}, \quad y=\frac{r_{+}-r_{-}}{2\kappa}, \qquad r_{\pm}=\sqrt{\rho^{2} + (z \pm \kappa)^{2}}.   \label{prolates}\ee

Is it worthwhile to mention that the above solution Eq.\ (\ref{ErnstKCH}) contains the real parameters $p$, $q$, $\gamma_{o}$, $\alpha$, $\beta$, and the half of the separation distance among sources $\kappa$, where the first three obey the constraints
\be  p^{2}+ q^{2}=1, \qquad |e^{-i\gamma_{o}}|=1.  \ee

Taking into account $y=1$ and $x=z/\kappa$, the Ernst potential on the upper part of the symmetry axis adopts the form
\bea \begin{split} {\cal{E}}(\rho=0,z)&=\frac{e_{+}(z)}{e_{-}(z)},\\
e_{\pm}(z)&=(p^{2}+\alpha^{2}-\beta^{2})z^{2} \mp 2\kappa[  (p+q\beta-ip\alpha)e^{-i\gamma_{o}}
\pm i(pq+\beta)]z\\
& +\kappa^{2}(p^{2}-\alpha^{2}+\beta^{2}+2i\alpha) \pm 2\kappa^{2}e^{-i\gamma_{o}} (q\alpha-ip \beta),\label{Ernstaxis}\end{split}\eea

\noi from which the first Geroch-Hansen multipolar moments \cite{Geroch, Hansen} can be explicitly computed once we apply the Fodor-Hoenselaers-Perj\'es procedure \cite{FHP}; they read \cite{MR}
\bea \begin{split}
M&=\frac{2\kappa(pP-pQ\alpha+qP\beta)}{p^{2}+\alpha^{2}-\beta^{2}},\qquad
J=M \left[\frac{(pq+\beta)M+\kappa(qQ\alpha+pP\beta)}{pP-pQ\alpha+qP\beta}-2J_{0} \right], \\
J_{0}&=-\frac{2\kappa(pQ+pP\alpha+qQ\beta)}{p^{2}+\alpha^{2}-\beta^{2}}, \qquad e^{-i\gamma_{o}}:=P-i Q, \label{ernstaxis}\end{split}\eea

\noi where $M$ and $J$ represent the total mass and total angular momentum of the system, respectively. Besides, $J_{0}$ is the NUT parameter.$^{[31]}$\footnotetext[31]{Ref.\ \cite{MR} does not consider the contribution of the NUT parameter $J_{0}$ inside the total angular momentum, it means that the full KCH metric contains two semi-infinite singularities located up and down along the symmetry axis.} Starting with the previous axis data, in Ref.\ \cite{MR} is provided the full KCH metric via the Sibgatullin method \cite{Sibgatullin}, which is written down in a closed analytical form by using the Perj\'es' factor structure \cite{Perjes}; it reads
\bea \begin{split} f&=\frac{N}{D}, \qquad
\omega=2J_{0}(y-1)+\frac{\kappa(y^{2}-1)F}{N},\qquad
e^{2\gamma}=\frac{N}{K_{0}^{2}(x^{2}-y^{2})^{4}}, \\
N&=\mu^{2}+(x^{2}-1)(y^{2}-1)\sigma^{2},\qquad D= N+ \mu \pi -(y^{2}-1)\sigma \tau, \qquad F=(x^{2}-1)\sigma \pi +\mu  \tau,\\
\mu&= p^{2}(x^{2}-1)^{2}+q^{2}(y^{2}-1)^{2}
+(\alpha^{2}-\beta^{2})(x^{2}-y^{2})^{2},\\
\sigma&=2\left[pq(x^{2}-y^{2})+\beta(x^{2}+y^{2})-2\alpha xy \right],\\
\pi&=(4/K_{0})\{K_{0}\left[ pPx(x^{2}+1)+2x^{2}+qQy(y^{2}+1)\right] +2(pQ+pP\alpha+qQ\beta)\\
&\times \left[ pqy(x^{2}-y^{2})+\beta y(x^{2}+y^{2})-2\alpha xy^{2}\right]-K_{0}(x^{2}-y^{2})\left[ (pQ \alpha-qP\beta)x+(qP\alpha-pQ\beta)y\right]\\
&-2(q^{2}\alpha^{2}+p^{2}\beta^{2})(x^{2}-y^{2})+4(pq+\beta)x(\beta x-\alpha y)\},\\
\tau&=(4/K_{0})\{K_{0}x\left[ (qQ\alpha+ pP\beta)(x^{2}-y^{2})+qP(y^{2}-1)\right] +(pQ+pP \alpha+qQ\beta)y\\
& \times \left[(p^{2}-\alpha^{2}+\beta^{2})(x^{2}-y^{2})+y^{2}-1 \right]-pQK_{0}y(x^{2}-1)-2p(q\alpha^{2}-q\beta^{2}-p \beta)(x^{2}-y^{2})\\
& +(pq+\beta)(y^{2}-1)
\},\\
K_{0}&=p^{2}+\alpha^{2}-\beta^{2}.
\label{MRmetric}\end{split}\eea

One should notice that the above metric is invariant under the change $\{p,q,P,Q,\alpha,\beta\}\rightarrow \{-p,-q,-P,-Q,\alpha,\beta\}$. Bearing in mind that asymptotically flat spacetimes can be obtained from Eq.\ (\ref{MRmetric}) when the NUT parameter $J_{0}$ is eliminated, there exist several possibilities to achieve such a task. On one hand, Yamazaki \cite{Yamazaki} proposed the solution
\be P= \frac{p+ q\beta}{\sqrt{(p+q\beta)^{2}+p^{2}\alpha^{2}}}, \qquad Q= -\frac{p \alpha}{\sqrt{(p+q\beta)^{2}+p^{2}\alpha^{2}}},\ee

\noi while on the other hand Manko and Ruiz \cite{MR} went beyond at the moment of considering the following solution
\be \alpha=-\frac{Q(p+q\beta)}{p P}. \label{MRcounter}\ee

Due to the fact that the metric function $\omega$ on the middle region among the sources ($x=1$, $y=z/\kappa$), acquires the form
\bea \begin{split} &q \alpha\left[Q K_{0}-(2p+P)\alpha\right] + \beta\left[pP K_{0}+(2p+P)q\beta-1+2p^{2}\right] -pq(1+pP)\\
&-(pQ+pP\alpha+qQ\beta)(q^{2}+\alpha^{2}-\beta^{2})=0, \end{split}\eea

\noi one notices that Yamazaki's approach does not simplify the above condition, while the second proposal considered by Manko and Ruiz factorizes it as follows
\be \left[(p^{2}-Q^{2})\beta^{2}-pq(1+pP+Q^{2})\beta-p^{2}(1+pP)\right]
\left[(p^{2}-Q^{2})\beta-pq(pP+Q^{2})\right]=0, \label{twofactors}\ee

\noi which eventually may lead us to the description of two-body systems of unequal co/counter-rotating BHs separated by a massless strut by choosing the first/second factor respectively. In this direction, over all the parametrization of \cite{MR}, the total mass $M$ and total angular momentum $J$ of the system were given in terms of dimensionless parameters $\{p,q,P,Q\}$, therefore, the analysis of the dynamics for such BH systems was mostly performed in a numerical way. For instance, the total mass $M$ and total angular momentum $J$ in the counter-rotating sector are obtainable from the second factor of Eq.\ (\ref{twofactors}) in combination with Eq.\ (\ref{MRcounter}); they assume the form
\be M=\frac{2\kappa (pP+q^{2})}{p^{2}-q^{2}}, \qquad J=\frac{2\kappa^{2}q\left[(1+2pP)^{2}-(p+P)^{2}\right]}{p(p^{2}-q^{2})^{2}}. \ee

To make matters worse, the situation is even much more complicated in the co-rotating sector, where after using the first factor of Eq.\ (\ref{twofactors}) together with Eq.\ (\ref{MRcounter}) one obtains
\bea \begin{split} M&=\frac{\kappa \left[\pm q\Delta_{o}-p(1+p^{2})-q^{2}P\right]}{p(p^{2}-q^{2})},\\
J&=\frac{\kappa M}{2p(p^{2}-q^{2})}\{\pm \Delta_{o}(2p^{2}P-2p-P)
+2q(1+p^{2}+pP)-qP(p^{2}-q^{2})(p-P)\}, \\
\Delta_{o}&:=\sqrt{4p^{2}(1+pP)+q^{2}(p+P)^{2}}. \end{split}\eea

This last point naturally motivates the present work to consider another more suitable parametrization which might invert the problem and establish a real physical representation of the KCH metric to describe interacting BHs in a more transparent form.

\vspace{-0.5cm}
\section{Extreme binary black holes in a physical representation}
\vspace{-0.5cm}
The problem of expressing the KCH metric with a more physical aspect can be tackled by adopting first a new representation for such a solution. In order to do so, we begin with a new suitable parametrization of the Ernst potential on the symmetry axis
\bea \begin{split}
&\hspace{-2.0cm} e(z)=\frac{z^{2}-[M + i(\mathfrak{q}+2J_{0})]z + \frac{2\Delta-R^{2}}{4} + \frac{\mathfrak{q}(P_{1} +P_{2})}{2M} -2 \mathfrak{q} J_{0}+ i\left(P_{1}- \frac{2J_{0}(P_{2}+M \mathfrak{q})}{\mathfrak{q}} \right) }{z^{2} + (M -i\mathfrak{q})z + \frac{2\Delta-R^{2}}{4} -\frac{\mathfrak{q}(P_{1} +P_{2})}{2M} + i P_{2}},\\
\Delta:=M^{2}-\mathfrak{q}^{2}, \label{ernstaxis}\end{split}\eea

\noi where the KCH solution contains now the five parameters $\{M, \mathfrak{q},R, P_{1}, P_{2}\}$  related to the set $\{p, q, \gamma_{o}, \alpha, \beta, \kappa\}$ via the expressions
\bea \begin{split}
\mathfrak{q}&=\frac{2\kappa[p(q+Q)+ p P \alpha +(1+qQ)\beta]}{p^{2}+\alpha^{2}-\beta^{2}}, \qquad R=2\kappa, \\ P_{1}&=\frac{2\kappa^{2}[(1-qQ)\alpha-pP\beta]}{p^{2}+\alpha^{2}-\beta^{2}}+
2\left(M+\frac{P_{2}}{\mathfrak{q}}\right)J_{0},\qquad
P_{2}= \frac{2\kappa^{2}[(1+qQ)\alpha+pP\beta]}{p^{2}+\alpha^{2}-\beta^{2}},\label{relations1}\end{split}\eea

\noi being $M$ and $J_{0}$ expressed lines up. The inverse relation among these two sets of parameters is completed if we construct once again the full KCH metric by using the Perj\'es' factor structure in the same way like in Ref.\ \cite{MR}, leading us to
\bea \begin{split}
p&=\frac{M(\alpha_{+}\beta_{+}-\alpha_{-}\beta_{-})}{\sqrt{(\alpha_{+}^{2}+M^{2}\beta_{-}^{2})
(\alpha_{-}^{2}+M^{2}\beta_{+}^{2})}}, \qquad
q=-\frac{\alpha_{+}\alpha_{-}+M^{2}\beta_{+}\beta_{-}}{\sqrt{(\alpha_{+}^{2}+M^{2}\beta_{-}^{2})
(\alpha_{-}^{2}+M^{2}\beta_{+}^{2})}},\\
P&=\frac{M(\alpha_{+}\beta_{+}+\alpha_{-}\beta_{-})}{\sqrt{(\alpha_{+}^{2}+M^{2}\beta_{-}^{2})
(\alpha_{-}^{2}+M^{2}\beta_{+}^{2})}},\qquad
Q=\frac{\alpha_{+}\alpha_{-}-M^{2}\beta_{+}\beta_{-}}{\sqrt{(\alpha_{+}^{2}+M^{2}\beta_{-}^{2})
(\alpha_{-}^{2}+M^{2}\beta_{+}^{2})}},\\
\alpha&=MR\left(\frac{[\mathfrak{q}\alpha_{+}+M(R + M)\beta_{-}][q_{o}- 2M^{2}R(\alpha_{-}-\mathfrak{q}\beta_{+})]- [\mathfrak{q}\alpha_{-}-M(R - M)\beta_{+}][q_{o}+ 2M^{2}R(\alpha_{+}-\mathfrak{q}\beta_{-})]}{(\alpha_{-}^{2}+M^{2}\beta_{+}^{2})
(\alpha_{+}^{2}+M^{2}\beta_{-}^{2})}\right),\\
\beta&=MR\left(\frac{[\mathfrak{q}\alpha_{+}+M(R + M)\beta_{-}][q_{o}- 2M^{2}R(\alpha_{-}-\mathfrak{q}\beta_{+})]+ [\mathfrak{q}\alpha_{-}-M(R - M)\beta_{+}][q_{o}+ 2M^{2}R(\alpha_{+}-\mathfrak{q}\beta_{-})]}{(\alpha_{-}^{2}+M^{2}\beta_{+}^{2})
(\alpha_{+}^{2}+M^{2}\beta_{-}^{2})}\right),\\
\kappa&=R/2, \qquad q_{o}:=\alpha_{+}\alpha_{-}+M^{2}\beta_{+}\beta_{-}, \qquad
\alpha_{\pm}:=M(\Delta \pm MR)-\mathfrak{q}(P_{1}+P_{2}), \qquad \beta_{\pm}:= 2P_{2}\pm \mathfrak{q}R.\label{relations2}\end{split}\eea

Additionally, the total angular momentum $J$ as well as the NUT charge $J_{0}$, in this new representation are reduced to
\bea \begin{split}
J&=M \mathfrak{q}-\frac{P_{1}-P_{2}}{2}+\frac{J_{0}P_{2}}{\mathfrak{q}}, \\ J_{0}&=\frac{\mathfrak{q}}{2M}\left(\frac{\mathfrak{q}^{2}(P_{1}+P_{2})^{2}-
M^{2}\left[4P_{1}P_{2}
-\Delta(R^{2}-\Delta)\right] }{\mathfrak{q}^{2}\left[M(R^{2}-\Delta)+\mathfrak{q}(2P_{1}+2P_{2}+M \mathfrak{q})\right]-M(2P_{2}+M \mathfrak{q})^{2}}\right). \label{Multipolarterms}\end{split}\eea

With the main purpose of describing the interaction among two extreme BHs separated by a conical singularity \cite{Bach,Israel}, the first equation that eliminates $J_{0}$ is
\be \mathfrak{q}^{2}(P_{1}+P_{2})^{2}-M^{2}\left[4P_{1}P_{2} -\Delta(R^{2}-\Delta)\right]=0 , \label{noNUT}\ee

\noi while after developing a few non-trivial calculations one gets a simple quadratic expression
for the axis condition $\omega(x=1,y=2z/R)=0$, that disconnects the region in between sources, namely
\bea \begin{split}
 \mathfrak{q}\left[M^{2}(R^{2}+MR+\mathfrak{q}^{2})(P_{1}-P_{2})^{2}
+(\Delta+MR)(\Delta-MR-R^{2})(P_{1}+P_{2})^{2}\right]&\\
-M^{2}(R^{2}-\Delta)
\left\{\left[M\mathfrak{q}^{2}+(R+M)(R^{2}+MR+\mathfrak{q}^{2})\right](P_{1}-P_{2})-
M\mathfrak{q}(R+M)(R^{2}-\Delta)\right\}=0,&
\label{conditionmiddle}\end{split}\eea

\noi and it is not complicated to show that Eqs.\ (\ref{noNUT}) and (\ref{conditionmiddle}) contain a trivial set of solutions, which explicitly are
\bea \begin{split}
P_{1,2}&= \frac{\mp \Delta[R^{2}+MR+\mathfrak{q}^{2}] + \epsilon M \sqrt{\Delta(R^{2}+MR+\mathfrak{q}^{2})^{2} + M^{2} \mathfrak{q}^{2}(R^{2}-\Delta)}}{2M\mathfrak{q}},\\
P_{1,2}&= \frac{ \mp  \mathfrak{q}\Delta + \epsilon M \sqrt{(R+M)^{2}(R^{2}-\Delta)+\mathfrak{q}^{2}\Delta}}{2(R+M)}, \qquad \epsilon=\pm 1,\\
 \label{solutions}  \end{split}\eea

\noi where the subindexes $1$ and $2$ are associated with $-$ and $+$ signs, while the sign of $\epsilon$ refers to the location of the sources. In what follows in this paper we are going to use $\epsilon=1$; this means that the first/second source will be located up/down, respectively. The aforementioned Eq.\ (\ref{solutions}) is giving us two $3$-parametric subfamilies of the KCH metric that we are going to explore in the next subsections. It is worth mentioning that Eq.\ (\ref{conditionmiddle}) has been derived recently in Ref.\ \cite{Cabrera} for the case of non-extreme sources. 

\vspace{-0.5cm}
\subsection{Co-rotating binary black holes}
\vspace{-0.5cm}
Using the first solution of Eq.\ (\ref{solutions}) it can be possible to describe a co-rotating two-body system of unequal Kerr sources separated by a massless strut, as a $3$-parametric subclass of the KCH metric. By means of Perjes's representation \cite{Perjes}, the Ernst potential ${\cal{E}}$ and the full metric are depicted by
\bea \begin{split} {\cal{E}}&=\frac{\Lambda-2\Gamma}{\Lambda+2\Gamma},\qquad f=\frac{N}{D}, \qquad
\omega=\frac{R(y^{2}-1)F}{2N},\qquad
e^{2\gamma}=\frac{N}{\mathfrak{q}^{4}R^{4}(x^{2}-y^{2})^{4}}, \\
\Lambda&=\mathfrak{q}^{2}(R^{2}-\Delta)(x^{2}-y^{2})^{2}+\Delta[\mathfrak{q}^{2}(x^{4}-1)+
(R+M)^{2}(y^{4}-1)]\\
&+2i\mathfrak{q}\left\{xy\left[\Delta(R+M)(x^{2}-y^{2})-M(MR+\Delta)(x^{2}+y^{2}-2)\right]-\delta_{1}
(x^{2}+y^{2}-2x^{2}y^{2})\right\},\\
\Gamma&=\left(\frac{\mathfrak{q}(\Delta-MR-R^{2})+i\delta_{1}}{MR[(R+M)^{2}+\mathfrak{q}^{2}]}\right)\{
\Delta\left[(R+M)^{2}+\mathfrak{q}^{2}\right]\left[ \mathfrak{q}x(x^{2}-1)-i(R+M)y(y^{2}-1)\right]\\
&-\mathfrak{q}\left\{M(\Delta+MR)\left[ (R+M)x-iqy\right]-\delta_{1}\left[ (R+M)y-i\mathfrak{q}x\right]\right\}(x^{2}-y^{2})\},\\
N&=\mu^{2}+(x^{2}-1)(y^{2}-1)\sigma^{2},\quad D= N+ \mu \pi-(1-y^{2})\sigma \tau, \quad F=(x^{2}-1)\sigma\pi-\mu \tau,\\
\mu&= \mathfrak{q}^{2}(R^{2}-\Delta)(x^{2}-y^{2})^{2}+\Delta\left[\mathfrak{q}^{2}(x^{2}-1)^{2}
+(R+M)^{2}(y^{2}-1)^{2}\right],\\
\sigma&=2\mathfrak{q}\left\{\mathfrak{q}^{2}Rx^{2}+\left[2M(\Delta+MR)-\mathfrak{q}^{2}R\right]y^{2}
-2\delta_{1}xy\right\},\\
\pi&=\left(\frac{4}{MR}\right)\{\mathfrak{q}^{2}x\left[M^{2}R\left(R(x^{2}-y^{2})+2Mx\right)+\Delta
(\Delta-MR-R^{2})(1+y^{2})-4M \delta_{1}y\right]\\
&+y\left[2M\left(\Delta(R+2M)(R^{2}+\mathfrak{q}^{2})+ M^{4}R \right)y+\delta_{1}\left( M(\Delta+MR)(1+y^{2})-\mathfrak{q}^{2}R(1+x^{2})\right) \right]\},\\
\tau&=\left(\frac{4 \mathfrak{q} \Delta}{MR}\right)\{-x
\left[(R^{2}+M R+\mathfrak{q}^{2})(Rx+2M)x+(R+M)(\Delta-MR-R^{2}) \right]\\
&+(1-x)y\left[M(R^{2}-\Delta)y+\delta_{1}(1+x)\right]+M\left((R+M)^{2}+\mathfrak{q}^{2}\right)\}\\
\delta_{1}&:=\epsilon \sqrt{\Delta(R^{2}+MR+\mathfrak{q}^{2})^{2} + M^{2} \mathfrak{q}^{2}(R^{2}-\Delta)}.
\label{extremecorotating}\end{split}\eea

It is feasible to prove from Eq.\ (\ref{relations2}) that the above metric is obtainable from the KCH metric \cite{KCH,MR} after doing the following changes in the real parameters:
\bea \begin{split}
p&=\frac{\mathfrak{q}}{\sqrt{(R+M)^{2}+\mathfrak{q}^{2}}}, \qquad q=-\frac{R+M}{\sqrt{(R+M)^{2}+\mathfrak{q}^{2}}}, \qquad e^{-i\gamma_{o}}=\frac{\mathfrak{q}(\Delta-MR-R^{2})+i\delta_{1}}{MR\sqrt{(R+M)^{2}
+\mathfrak{q}^{2}}},\\
\alpha&=\frac{\mathfrak{q}\delta_{1}}{\Delta\left[(R+M)^{2}+\mathfrak{q}^{2}\right]},\qquad
\beta=\frac{M\mathfrak{q} (\Delta+MR)}{\Delta\left[(R+M)^{2}+\mathfrak{q}^{2}\right]}.
\end{split}\eea

On the other hand, the Komar integrals \cite{Komar} for each mass and angular momentum can be calculated through the Tomimatsu's formulae \cite{Tomimatsu0}:
\be  M_{i}=-\frac{1}{8\pi}\int_{H_{i}} \omega\, {\rm{Im}}({\cal{E}}_{z}) d\varphi dz, \qquad J_{i}=-\frac{1}{8\pi}\int_{H_{i}}\omega\, \left(1+\frac{1}{2}\omega\, {\rm{Im}}({\cal{E}}_{z}) \right) d\varphi dz,\label{Tomi} \ee

\noi where the integrals must be evaluated over the corresponding horizon $H_{i}$. Apparently it seems quite complicated to develop such a goal, nevertheless, the technical difficulty of finding the correct formulas for the Komar masses and angular momenta of the sources can be circumvented by taking into account a limit process after expanding the above expressions around the values taken by $x$ and $y$ on the regions surrounding to each BH; for instance, if we are surrounding the upper BH, one can take into the computer code $x=1+\varepsilon, y=1$ in the region on the axis $z>R/2$, for $|\varepsilon|<<1$, but in the region in between $|z|<R/2$ we put now $x=1, y=1-\varepsilon$. A trivial calculation yields the expressions
\bea \begin{split} M_{1,2}&=\frac{M}{2} \mp \frac{\delta_{1}}{2(\Delta+M R)},\\
J_{1,2}&= M_{1,2}\left[\frac{\mathfrak{q}}{2}-\frac{\Delta(R+M)(R^{2}+MR+\mathfrak{q}^{2})\pm (\Delta+MR)\delta_{1}}
{2M\mathfrak{q}(R^{2}-\Delta)} \right],\label{massesco} \end{split}\eea

\noi and it is easy to observe that $M=M_{1}+M_{2}$. Furthermore, the expression $J=J_{1}+J_{2}$ allows us to recover the aforementioned  Eq.\ (\ref{Multipolarterms}) for the total angular momentum in the absence of the NUT charge, namely
\be J=M\mathfrak{q}-\frac{P_{1}-P_{2}}{2}=M\mathfrak{\mathfrak{q}}+ \frac{\Delta (R^{2}+MR+\mathfrak{q}^{2})}{2M\mathfrak{q}}, \label{momenta1}\ee

\noi whereas the difference between the values of the masses yields the relation
\be M_{2}-M_{1}=\frac{\delta_{1}}{\Delta+M R},\ee

\noi and eventually one arrives to a bicubic equation for solving
\bea \begin{split} &\mathfrak{q}^{6}+3a_{1}\mathfrak{q}^{4}+3a_{2}\mathfrak{q}^{2}+a_{3}=0,\\
a_{1}&:=(1/3)[2R^{2}+2MR-2M^{2}+(M_{1}-M_{2})^{2}], \\
a_{2}&:=(1/3)(R+M)[(R-M)(R^{2}+2MR-M^{2})- 2M(M_{1}-M_{2})^{2}],\\
a_{3}&:=-M^{2}(R+M)^{2}\left[R^{2}-(M_{1}-M_{2})^{2}\right], \label{bicubic1} \end{split}\eea

\noi whose explicit roots are given by
\bea \begin{split}
\mathfrak{q}^{2}_{(k)}&= -a_{1} + e^{i 2\pi k/3}\left[b_{o}+ \sqrt{b_{o}^{2}-a_{o}^{3}}\right]^{1/3}+ e^{-i 2\pi k/3}a_{o} \left[b_{o}+ \sqrt{b_{o}^{2}-a_{o}^{3}}\right]^{-1/3},\\
a_{o}&:=a_{1}^{2}-a_{2},\qquad
b_{o}:=(1/2)\left[3a_{1}a_{2}-a_{3}-2a_{1}^{3}\right], \quad k=0,1,2. \label{theq} \end{split}\eea

In this particular case we choose $k=0$ since it defines entirely a real parameter $\mathfrak{q}$ which starts and ends at the same value given by the total mass $M$, where the coordinate distance runs from $R=0$ until $R \rightarrow \infty$. The substitution of this real solution into Eq.\ (\ref{massesco}) permits us to demonstrate that during the merging process ($R=0$) each individual angular momentum $J_{i}$ is related to its corresponding mass $M_{i}$ by means of \cite{Cabrera}
\be \frac{J_{1}}{M_{1}^{2}}=1+\frac{M_{2}}{M_{1}}, \qquad \frac{J_{2}}{M_{2}^{2}}=1+\frac{M_{1}}{M_{2}}, \label{fusion}\ee

\noi where such a process conceives a single extreme BH of mass $M=M_{1}+M_{2}$ and angular momentum $J=J_{1}+J_{2}$, satisfying exactly a well-known formula for extreme BHs \cite{Cabrera}
 \be J=J_{1}+J_{2}=(M_{1}+M_{2})^{2}.\ee

Moreover, when the sources are far away from each other, in the limit $R \rightarrow \infty$ are recovered the simple expressions for extreme BHs, namely
\be \frac{J_{1}}{M_{1}^{2}}=1, \qquad \frac{J_{2}}{M_{2}^{2}}=1. \label{isolatedcorotating}\ee

\begin{figure}[ht]
\begin{minipage}{0.49\linewidth}
\centering
\includegraphics[width=7.5cm,height=5.8cm]{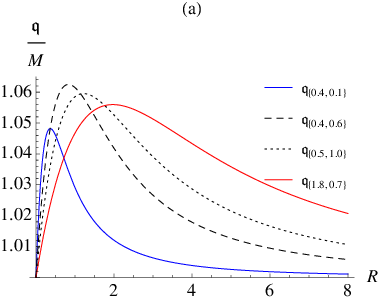}
\end{minipage}
\begin{minipage}{0.49\linewidth}
\centering
\includegraphics[width=7.5cm,height=5.8cm]{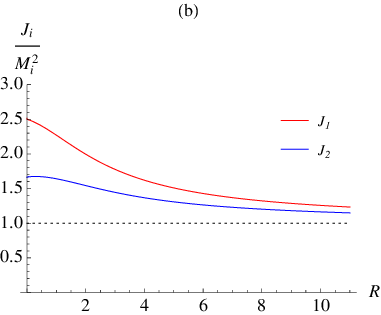}
\end{minipage}
\caption{(a) Behavior for the parameter $\mathfrak{q}$ in the co-rotating case taking different values for the masses $M_{1}$ and $M_{2}$ denoted by the subscripts inside the brackets, respectively. (b) The angular momenta $J_{1}$ and $J_{2}$ for the values $M_{1}=1.2$ and $M_{2}=0.8$.}
\label{qJcoro}\end{figure}

All these features already mentioned can be noticed in Fig.\ \ref{qJcoro}. Regarding now the dynamical aspects of this co-rotating two-body system, the interaction force associated with the strut can be computed straightforwardly by using the formula \cite{Israel,Weinstein}, to obtain
\be \mathcal{F}=\frac{1}{4}(e^{-\gamma_{s}}-1)=\frac{\Delta\left[\mathfrak{q}^{2}-(R+M)^{2}\right]}
{4\left(\Delta+M R\right)^{2}}\equiv\frac{M_{1}M_{2}[(R+M)^{2}-\mathfrak{q}^{2}]}
{(R^{2}-\Delta)[(R+M)^{2}+\mathfrak{q}^{2}]},
\label{forceco} \ee

\noi where $\gamma_{s}$ is the value of the metric function $\gamma$  evaluated on the region of the conical singularity; i.e., $\gamma(x=1)$. The strut prevents the BHs from falling onto each other; it means that as both horizons are getting closer and closer, the interaction force $\mathcal{F}\rightarrow \infty$. The minimal distance occurs when $R\rightarrow 0$, and for that case $\mathfrak{q}\rightarrow M$ [see Eq.\ (\ref{bicubic1}) or  Fig.\ \ref{qJcoro} (a)]. Let us now assume that the sources move away from each other, thus, in the limit $R\rightarrow \infty$ one gets the following expansion:
\be \mathcal{F}\simeq\frac{M_{1}M_{2}}{R^{2}}\left[1-\frac{2M^{2}}{R^{2}}+\frac{4M(M_{1}^{2}
+8M_{1}M_{2}+M_{2}^{2})}{R^{3}}+
O\left(\frac{1}{R^{4}}\right)\right],\ee

\noi that matches with the formula already given by Dietz and Hoenselaers \cite{DH} once we put the condition for extreme co-rotating sources given by Eq.\ (\ref{isolatedcorotating}). The strut might be removed if we consider $|\mathfrak{q}|=R+M$ in the above formula Eq.\ (\ref{forceco}). Nevertheless, as was demonstrated first by Hoenselaers \cite{Hoenselaers}, an absence of a strut might produce the appearance of naked singularities (ring singularities) off the axis, since at least one of the two masses will be negative even yet if the total mass of the system does not violate the positive mass theorem \cite{SchoenYau1, SchoenYau2}. The last statement can be confirmed directly from Eq.\ (\ref{massesco}).

To conclude the subsection, the identical case $M_{1}=M_{2}=m$,\, $J_{1}=J_{2}=j$ is recovered when is imposed the condition $\delta_{1}=0$, and is also taken into account a simple redefinition $\mathfrak{q}\rightarrow 2\mathfrak{q}$, thus, one arrives to the extreme condition for identical co-rotating BHs, which was considered earlier in \cite{Costa, CCLP}
\be m^{2}-\mathfrak{q}^{2}\left(1-\frac{4m^{2} (R^{2}-4m^{2}+4\mathfrak{q}^{2})}{[R(R+2m)+4\mathfrak{q}^{2}]^{2}}\right)=0,\ee

\noi where it can be shown that such a condition for identical extreme co-rotating BHs leads us to a bicubic equation, which is precisely the identical case of Eq.\ (\ref{bicubic1}). Furthermore, after replacing such a condition in Eq.\ (\ref{momenta1}) [or Eq.\ (\ref{massesco})], the final expression for the equal angular momentum acquires the form \cite{CCLP}
\be j= \frac{m\mathfrak{q}[(R+2m)^{2}+4\mathfrak{q}^{2}]}{R(R+2m)+4\mathfrak{q}^{2}}.\ee

For identical constituents, the values for the angular momentum $j$ are contained within the interval $1<j/m^{2}\leq 2$ \cite{Costa}, while for nonequal sources the ratio $J_{i}/M_{i}^{2}$ can be greater or lower than $2$ [see Eq.\ (\ref{fusion}) or  Fig.\ \ref{qJcoro} (b)]. This peculiarity was first pointed in Ref.\ \cite{MR} using numerical arguments.

\vspace{-0.3cm}

\vspace{-1.1cm}
\subsection{Counter-rotating binary black holes}
\vspace{-0.5cm}
Regarding the second solution of Eq.\ (\ref{solutions}) which is referring to counter-rotating binary systems of unequal Kerr BHs also apart by a strut, where now the $3$-parametric member of the KCH exact solution is represented as follows:
\bea \begin{split} {\cal{E}}&=\frac{\Lambda-2\Gamma}{\Lambda+2\Gamma},\qquad f=\frac{N}{D}, \qquad
\omega=\frac{R(y^{2}-1)F}{2N},\qquad
e^{2\gamma}=\frac{N}{R^{6}(R+M)^{4}(x^{2}-y^{2})^{4}}, \\
\Lambda&=R\{(R+M)^{2}\left[(R^{2}-\Delta)(x^{2}-y^{2})^{2}+\Delta(x^{4}-1)\right] +\mathfrak{q}^{2}\Delta(y^{4}-1)\\
&+2i(R+M)\left(\mathfrak{q}xy\left[2\Delta(y^{2}-1)-R(R+M)(x^{2}+y^{2}-2)\right] -M\delta_{2}(x^{2}+y^{2}-2x^{2}y^{2})\right) \},\\
\Gamma&=\left(\frac{\Delta+M R+i\delta_{2}}{(R+M)^{2}+\mathfrak{q}^{2}}\right)\{\Delta
\left[(R+M)^{2}+\mathfrak{q}^{2}\right]\left[ (R+M)x(x^{2}-1)+i\mathfrak{q}y(y^{2}-1)\right] \\&+(R+M)\left\{\mathfrak{q}(R^{2}+MR-\Delta)\left[\mathfrak{q} x+i(R+M)y\right]-M\delta_{2}\left[\mathfrak{q} y+i(R+M)x\right]\right\}(x^{2}-y^{2})\}, \\
N&= \mu^{2}+(x^{2}-1)(y^{2}-1)\sigma^{2},\quad D= N+ \mu \pi-(1-y^{2})\sigma \tau, \quad F=(x^{2}-1)\sigma\pi-\mu \tau,\\
\mu&= R\left\{(R+M)^{2}\left[(R^{2}-\Delta)(x^{2}-y^{2})^{2}+\Delta(x^{2}-1)^{2}\right]  +\mathfrak{q}^{2}\Delta(y^{2}-1)^{2} \right\},\\
\sigma&=2R(R+M)\left[\mathfrak{q} R(R+M)(x^{2}+y^{2})-2\mathfrak{q}\Delta y^{2}-2M\delta_{2}xy\right],\\
\pi&=(4/R)\{ R(R+M)x\left[MR(R+M)\left(R(x^{2}-y^{2})+2Mx\right)+\Delta(MR+\Delta)(1+y^{2})\right]+ \mathfrak{q}Ry\\
&\times\left\{2\mathfrak{q}y\left[R(R+M)^{2}-\Delta(R+2M)\right]-\delta_{2}
\left[(R+M)\left(R(x^{2}-y^{2})+4Mx\right)+\Delta(1+y^{2})\right]\right\}\},\\
\tau&=4\Delta\{ \mathfrak{q} \left\{(R+M)^{2}+\mathfrak{q}^{2}+ x\left[MR+\Delta-(R+M)x(Rx+2M)\right] +(R^{2}-\Delta)(x-1)y^{2}\right\}\\
&- \delta_{2}(R+M)y(x^{2}-1)\}, \qquad \delta_{2}:=\epsilon\sqrt{(R+M)^{2}(R^{2}-\Delta)+\mathfrak{q}^{2}\Delta} ,
 \label{extremecounter}\end{split}\eea

\noi and this particular metric can be developed from the KCH metric \cite{KCH, MR} by means of
\bea \begin{split}
p&=\frac{R+M}{\sqrt{(R+M)^{2}+\mathfrak{q}^{2}}}, \qquad q=\frac{\mathfrak{q}}{\sqrt{(R+M)^{2}+\mathfrak{q}^{2}}}, \qquad e^{-i\gamma_{o}}=\frac{\Delta+MR+i\delta_{2}}{R\sqrt{(R+M)^{2}
+\mathfrak{q}^{2}}},\\
\alpha&=\frac{M(R+M)\delta_{2}}{\Delta\left[(R+M)^{2}+\mathfrak{q}^{2}\right]},\qquad
\beta=\frac{\mathfrak{q} (R+M) (R^{2}+MR-\Delta)}{\Delta\left[(R+M)^{2}+\mathfrak{q}^{2}\right]}.
\end{split}\eea

\noi where we have substituted the second solution of Eq.\ (\ref{solutions}) inside Eq.\ (\ref{relations2}). The corresponding masses $M_{i}$ and angular momenta $J$ are given, respectively, by
\bea \begin{split} M_{1,2}&=\frac{M}{2} \mp \frac{\mathfrak{q}( R^{2}+MR-\Delta)}{2 \delta_{2}},\\
J_{1,2}&= M_{1,2}\left[\frac{\mathfrak{q}}{2} \left(2-\frac{R^{2}}{R^{2}-\Delta}\right) \mp \frac{(\Delta+MR)\delta_{2}}{2(R+M)
(R^{2}-\Delta)}\right],
\label{massescounter} \end{split}\eea

\noi where once again we have that $M=M_{1}+M_{2}$. The expression of the total angular momentum $J=J_{1}+J_{2}$ agrees with Eq.\ (\ref{Multipolarterms}), acquiring the final form
\be J=M\mathfrak{q}-\frac{P_{1}-P_{2}}{2}= \mathfrak{q}\left(M+ \frac{\Delta }{2(R+M)}\right), \label{momenta2}\ee

\noi but now the difference between both masses is giving us
\be M_{2}-M_{1}= \frac{\mathfrak{q}(R^{2}+MR-\Delta)}{\delta_{2}},\ee

\noi yielding to another bicubic equation
\bea \begin{split} &\mathfrak{q}^{6}+3b_{1}\mathfrak{q}^{4}+3b_{2}\mathfrak{q}^{2}+b_{3}=0,\\
b_{1}&:=(1/3)[2R^{2}+2MR-2M^{2}+(M_{1}-M_{2})^{2}], \\
b_{2}&:=(1/3)[(R^{2}+MR-M^{2})^{2}-(M_{1}-M_{2})^{2}(R^{2}+2MR+2M^{2})],\\
b_{3}&:=-(M_{1}-M_{2})^{2}(R+M)^{2}\left(R^{2}-M^{2}\right), \label{bicubic} \end{split}\eea

\noi which is having the roots
\bea \begin{split}
\mathfrak{q}^{2}_{(k)}&= -b_{1} + e^{i 2\pi k/3}\left[b_{o}+ \sqrt{b_{o}^{2}-a_{o}^{3}}\right]^{1/3}+ e^{-i 2\pi k/3}a_{o} \left[b_{o}+ \sqrt{b_{o}^{2}-a_{o}^{3}}\right]^{-1/3},\\
a_{o}&:=b_{1}^{2}-b_{2},\qquad
b_{o}:=(1/2)\left[3b_{1}b_{2}-b_{3}-2b_{1}^{3}\right], \quad k=0,1,2. \label{theqcounter} \end{split}\eea

Let us also consider the interaction force among the BHs, where now contains the following aspect
\be \mathcal{F}=\frac{\Delta [(R+M)^{2}-\mathfrak{q}^{2}]}{4[(R+M)^{2}(R^{2}-\Delta)+\mathfrak{q}^{2}\Delta]}
\equiv\frac{M_{1}M_{2}[(R+M)^{2}-\mathfrak{q}^{2}]}{(R^{2}-\Delta)[(R+M)^{2}+\mathfrak{q}^{2}]}.
\label{forcecounter} \ee

\noi Therefore, the expression of the force assumes an equivalent final form in both co/counter-rotating configurations of interacting BHs, but their dynamical and thermodynamical characteristics will differ considerably each other at the moment of choosing values for $\mathfrak{q}$ that satisfy the cubic equation in each sector. The well-known identical counter-rotating BH systems are achieved by setting $\mathfrak{q}=0$ in the above formulas of this subsection, from where one gets $M_{1}=M_{2}=m$ and $J_{1}=-J_{2}=-j$. Such configurations where described first analytically by Varzugin \cite{Varzugin} after solving the corresponding Riemann-Hilbert problem, later on, Herdeiro \emph{et al.} \cite{Herdeiro} provided several dynamical and thermodynamical aspects for these binary systems; in particular, they recognized the limit value $R\rightarrow 2m$ in which the merging process befalls, and the relation $|j|>m^{2}$ that violates the Kerr bound. In addition, Manko \emph{et al.} \cite{MR,MRRS} identified clearly the $2$-parametric subfamily member of the KCH metric which is recovered after settling $\mathfrak{q}=0$ in Eq.\ (\ref{extremecounter}). Last but not least, Tomimatsu's equilibrium configurations without a supporting strut can be achieved whether $\mathfrak{q}=R+M$, and $M=R(l-1)/(2-l)$ \cite{Tomimatsu0,Hoenselaers}.

Continuing with the description and excluding the identical case, where now $\mathfrak{q} \neq 0$, we have noticed at least two possibilities in the relations between the masses given by the phase $k=0$ in Eq.\ (\ref{theqcounter}), where the coordinate distance $R$ is running from $R=M_{1}+M_{2}$ until $R \rightarrow \infty$. Without loss of generality, let us suppose that $M_{2}>M_{1}$, in this regard $\mathfrak{q}$ acquires the final value $\mathfrak{q}= M_{2}-M_{1}$ at infinity, while its initial value depends on which is the ratio between the masses at the moment that both sources are getting closer each other. On one hand if $M_{2}/M_{1}<(3 + \sqrt{5})/2 \simeq 2.61803 $ the real parameter $\mathfrak{q}$ tends to a value closely to zero, but is never touching it!, then $ \mathcal{F} \rightarrow \infty$ as $M_{2}$ approaches the value of $M_{1}$. On the other hand, if $M_{2}/M_{1}>(3+\sqrt{5})/2$, $\mathfrak{q}$ takes an initial value given by
\be \mathfrak{q}= \sqrt{\left(\frac{M_{2}-M_{1}}{2}\right)\left[ \sqrt{25M^{2}-4M_{1}M_{2}}-(M_{2}-M_{1})\right]-M^{2}},\ee

\noi but now the force remains finite. Fixing the mass $M_{1}=1$ and taking different values for the mass $M_{2}$ and the coordinate distance $R$, in Table \ref{table1} is provided several values for $\mathfrak{q}$, the angular momenta, and the force during the merging process. Some of these values are depicted below in Fig.\
\ref{Counterbodies}. Finally, when the sources are far away, the force behaves as
\be \mathcal{F}\simeq\frac{M_{1}M_{2}}{R^{2}}\left[1-\frac{2(M_{1}^{2}-4M_{1}M_{2}+M_{2}^{2})}{R^{2}}+
\frac{4(M_{1}-M_{2})^{2}M}{R^{3}}+O\left(\frac{1}{R^{4}}\right)\right],\ee

\noi and thereby, such result matches one more time with the expression of \cite{DH} due to the fact that the individual angular momenta and masses satisfy the following relations at infinity
\be \frac{J_{1}}{M_{1}^{2}}=-1, \qquad \frac{J_{2}}{M_{2}^{2}}=1,  \qquad J_{1}<0, \, J_{2}>0. \ee

\vspace{-0.8cm}
\begin{figure}[ht]
\begin{minipage}{0.49\linewidth}
\centering
\includegraphics[width=7.5cm,height=6.0cm]{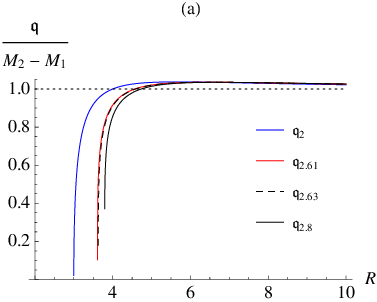}
\end{minipage}
\begin{minipage}{0.49\linewidth}
\centering
\includegraphics[width=7.5cm,height=6.0cm]{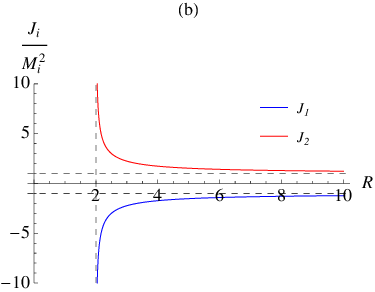}
\end{minipage}\vspace{0.5cm}
\begin{minipage}{0.49\linewidth}
\centering
\includegraphics[width=7.5cm,height=6.0cm]{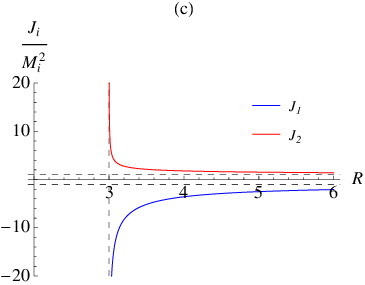}
\end{minipage}
\begin{minipage}{0.49\linewidth}
\centering
\includegraphics[width=7.5cm,height=6.0cm]{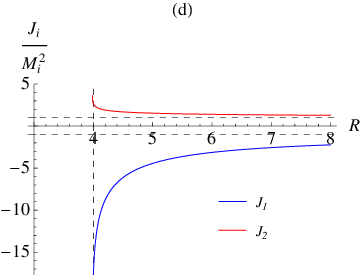}
\end{minipage}
\caption{(a) The parameter $\mathfrak{q}$ for counter-rotating BH systems fixing $M_{1}=1$ and assigning several values in the mass $M_{2}$ labeled by the subscripts. The angular momenta $J_{1}$ and $J_{2}$ for different values of $M_{2}$, where the merging limit is indicated by a vertical line given at the distance $R=M_{1}+M_{2}$; for (b) $M_{2}=1$, (c) $ M_{2}=2$, (d) $M_{2}=3$.}
\label{Counterbodies}\end{figure} \npg

\vspace{-0.3cm}
\begin{table}[ht]
\centering
\begin{tabular}{c c c c c c c c  }
\hline \hline
$M_{1}$&$ M_{2}$& $R$ & $ \mathfrak{q}$ & $J_{1}$ & $J_{2}$ & $J$ & $\mathcal{F}$  \\ \hline
  1 & 1 & 2  & 0 & -$\infty$  & $\infty$    & 0 & $\infty$    \\
  1 & 2 & 3.0001  & 0.0245 & -367.439  & 367.531    & 0.0918 & 1667.17   \\
  1 & 2.618 & 3.6181  & 0.2411 & -86.6621  & 87.7516 & 1.0895 & 44.3816\\
  1 & 2.62 & 3.6201 & 0.2498  & -83.7756 & 84.9047   & 1.1291   & 41.4199    \\
  1 & 3 & 4.0001 & 1.3038 & -17.679 & 24.0594  & 6.3804 & 1.6727 \\
  \hline \hline
\end{tabular}
\caption{Some numerical values for extreme counter-rotating BHs. The most violent merging process occurs at the limit value $R=2m$ and it corresponds to the case of identical sources $M_{1}=M_{2}=m$, on which the interaction force $ \mathcal{F} = \infty$ and each identical angular momentum $ |j|=\infty$, in agreement with Ref.\ \cite{Herdeiro}.}
\label{table1}
\end{table}

\vspace{-0.5cm}
\section{Concluding remarks}
\vspace{-0.5cm}
In the present paper we have worked out a concise physical representation for the two asymptotically flat
$3$-parametric subfamilies of the KCH metric \cite{KCH}, that may be useful to describe in a more transparent way the interactions between co/counter rotating binary BHs separated by a massless strut. In our opinion, this new parametrization is more suitable than the one presented in \cite{MR}, when we want to describe the dynamical and physical properties of extreme binary systems; in particular, at the moment of choosing values on the masses and the separation distance. Additionally, our analysis has revealed that both descriptions of co/counter-rotating binary configurations are contained within the same formula for the interaction force, but their dynamical aspects differ from each other after solving a proper bicubic equation in each sector. These bicubic equations can be understood as dynamical laws for interacting BHs with struts and are special cases of that one obtained previously in Ref.\ \cite{Cabrera}; it reads
\bea \begin{split} &\mathfrak{q}^{3}-(a_{1}+a_{2})\mathfrak{q}^{2}+(R+M)^{2}\mathfrak{q}
-(R+M)[a_{1}(R+M_{1}-M_{2})+a_{2}(R-M_{1}+M_{2})]=0,\label{condition}\end{split}\eea

\noi being $a_{i}\equiv J_{i}/M_{i}$, \, $i=1,2,$ the angular momentum per unit mass. So, once we substitute the Komar parameters $J_{i}$ of both co/counter-rotating two-body systems, their corresponding bicubic equations will emerge. Finally, we would like to pointed out that our physical representation of the KCH metric leads us to show clearly that the extreme solution saturates the Gabach Clement inequality \cite{Maria}
\be \sqrt{1 + 4 \mathcal{F}} = \frac{8\pi |J_{i}|}{S_{i}},\ee

\noi where $\mathcal{F}$ is given by Eqs.\ (\ref{forceco}) and (\ref{forcecounter}), while $S_{i}$ represents the area of the horizon $S$ in the extreme limit case, obtainable after establishing $\sigma_{i}=0$ in the expression (36) of Ref.\ \cite{Cabrera}, having that

\be S_{i}=4\pi \frac{M_{i}^{2}[(R+M)^{2}+\mathfrak{q}^{2}-2a_{i}\mathfrak{q}]^{2}+
a_{i}^{2}(R^{2}-\Delta)^{2}}{R^{2}[(R+M)^{2}+\mathfrak{q}^{2}]},\ee

\noi and therefore, it can be shown that the equality is reached after placing the angular momenta on each rotating sector.

\vspace{-0.5cm}
\section*{Acknowledgments}
\vspace{-0.5cm}
This work was supported by PRODEP, M\'exico, grant no 511-6/17-7605 (UACJ-PTC-367).


\begin{thebibliography}{99}
\bibitem{KCH}{W. Kinnersley and D. M. Chitre, J. Math. Phys. (N.Y.) \textbf{19}, 2037
(1978).}

\bibitem{KramerNeugebauer}{D. Kramer and G. Neugebauer, Phys. Lett. A \textbf{75}, 259
(1980).}

\bibitem{Yamazaki}{M. Yamazaki, Prog. Theor. Phys. \textbf{63}, 1950 (1980).}

\bibitem{NUT}{E. Newman, L. Tamburino, and T. Unti, J. Math. Phys. (N.Y.) \textbf{4}, 915
(1963).}

\bibitem{TS}{A. Tomimatsu and H. Sato, Prog. Theor. Phys. \textbf{50}, 95 (1973).}

\bibitem{MR}{V. S. Manko and E. Ruiz, Prog. Theor. Phys. \textbf{125}, 1241 (2011).}

\bibitem{Bach}{R. Bach and H. Weyl, Math. Z. \textbf{13}, 134 (1922).}

\bibitem{Israel}{W. Israel, Phys. Rev. D \textbf{15}, 935 (1977).}

\bibitem{Komar}{A. Komar, Phys. Rev. \textbf{113}, 934 (1959).}

\bibitem{Papapetrou}{A. Papapetrou, Proc. R. Irish Acad. Sect. A  \textbf{51}, 191 (1947).}

\bibitem{Ernst}{F. J. Ernst, Phys. Rev. \textbf{167}, 1175 (1968).}

\bibitem{Geroch}{R. Geroch, J. Math. Phys. \textbf{11}, 2580 (1970).}

\bibitem{Hansen}{R. O. Hansen,  J. Math. Phys. \textbf{15}, 46 (1974).}

\bibitem{FHP}{D. Fodor , C. Hoenselaers, and Z. Perj\'es, J. Math. Phys. \textbf{30}, 2252
(1989).}

\bibitem{Sibgatullin}{N. R. Sibgatullin, Oscillations and Waves in Strong Gravitational and
Electromagnetic Fields, Springer-Verlag, Berlin, 1991.}

\bibitem{Perjes}{Z. Perj\'es, J. Math. Phys. (N.Y.) \textbf{30}, 2197 (1989).}

\bibitem{Cabrera}{I. Cabrera-Munguia, arXiv:1806.05442v1.}

\bibitem{Tomimatsu0}{A. Tomimatsu, Prog. Theor. Phys. \textbf{70}, 385 (1983).}

\bibitem{Weinstein}{Weinstein G, Commun. Pure Appl. Math. \textbf{43}, 903 (1990).}

\bibitem{DH}{W. Dietz C. Hoenselaers,  Ann. Phys. (N.Y.) \textbf{165}, 319 (1985).}

\bibitem{Hoenselaers}{C. Hoenselaers, Prog. Theor. Phys. \textbf{72}, 761 (1984).}

\bibitem{SchoenYau1}{R. Schoen and S.-T. Yau, Commun. Math. Phys. \textbf{65}, 45 (1979). }

\bibitem{SchoenYau2}{R. Schoen and S.-T. Yau, Commun. Math. Phys. \textbf{79}, 231 (1981). }

\bibitem{Costa}{Miguel S. Costa, Carlos A. R. Herdeiro, and Carmen Rebelo, Phys. Rev. D,
\textbf{79}, 123508 (2009).}

\bibitem{CCLP}{I. Cabrera-Munguia, V. E. Ceron, L. A. L\'opez, and Omar Pedraza, Phys. Lett.
B \textbf{772}, 10 (2017).}

\bibitem{Varzugin}{G. G. Varzugin, Theor. Math. Phys. \textbf{116}, 1024 (1998).}

\bibitem{Herdeiro}{C. A. R. Herdeiro and C. Rebelo, JHEP \textbf{0810}, 017 (2008).}

\bibitem{MRRS}{V. S. Manko, E. D. Rodchenko, E. Ruiz, and B. I. Sadovnikov, Phys. Rev. D
\textbf{78}, 123508 (2008).}

\bibitem{Maria}{M. E. Gabach Clement, Class. Quantum Grav. \textbf{29}, 165008 (2012).}

\end{thebibliography}
\end{document}